\documentclass[manuscript]{aastex}

\shorttitle{Emitting electrons spectra of Mrk 421}
\shortauthors{Yan \& Zhang et al.}

\begin{document}


\title{Emitting electrons spectra and acceleration processes in the jet of Mrk 421: from low state to giant flare state}


\author{Dahai Yan\altaffilmark{1}, Li Zhang\altaffilmark{1}, Qiang Yuan\altaffilmark{2}, Zhonghui Fan\altaffilmark{1}, Houdun Zeng\altaffilmark{1}}


\altaffiltext{1}{Department of Physics, Yunnan University, Kunming
650091, Yunnan, China} \altaffiltext{2}{Key Laboratory of Particle
Astrophysics, Institute of High Energy Physics, Chinese Academy of
Sciences, Beijing 100049, China}
 \email{lizhang@ynu.edu.cn}


\begin{abstract}
We investigate the electron energy distributions (EEDs) and the
acceleration processes in the jet of Mrk 421 through fitting the
spectral energy distributions (SEDs) in different active states
in the frame of a one-zone synchrotron self-Compton (SSC) model.
After assuming two possible EEDs formed in different acceleration
models: the shock accelerated power-law with exponential cut-off
(PLC) EED and the stochastic turbulence accelerated log-parabolic
(LP) EED, we fit the observed SEDs of Mrk 421 in both low and
giant flare states by using the Markov Chain Monte Carlo
(MCMC) method which constrains the model parameters in a more
efficient way. Our calculating results indicate that (1) the PLC
and LP models give comparably good fits for the SED in low state,
but the variations of model parameters from low state to flaring
can be reasonably explained only in the case of the PLC in low state; and (2)
the LP model gives better fits compared to the PLC model for the
SED in flare state, and the intra-day/night variability
observed at GeV-TeV bands can be accommodated only in the LP model.
The giant flare may be attributed to the stochastic turbulence
re-acceleration of the shock accelerated electrons in low state.
Therefore, we may conclude that shock acceleration is dominant in
low state, while stochastic turbulence acceleration is dominant in
flare state. Moreover, our result shows that the extrapolated TeV
spectra from the best-fit SEDs from optical through GeV with the
two EEDs are different. It should be considered in caution when
such extrapolated TeV spectra are used to constrain extragalactic
background light (EBL) models.
\end{abstract}



\section{Introduction}

Blazars are the most extreme class of active galactic nucleies (AGNs).
Their spectral energy distributions (SEDs) are characterized by
two distinct bumps. The first bump, which is located at the low energy
band from
radio through UV or X-rays,
is generally explained by the synchrotron emission from relativistic
electrons in a jet
that is closely aligned to the line of sight.
The second bump, which is located
at the high-energy band, could be produced by inverse Compton
(IC) scattering of the relativistic electrons \citep[the so-called
leptonic model; e.g.,][]{bott07}.
The seed photons for IC process
can be synchrotron photons \citep[synchrotron self-Compton, SSC;][]{rees,Maraschi92}
or external
radiation fields \citep[EC;][]{dermer93,Sikora94}. The hadronic model
is an alternative explanation for the high energy emissions
from blazars \citep[e.g.,][]{Mannheim,Mucke,hadron12,dermer12}.

In the leptonic model, a power-law EED with an exponential
high-energy cutoff (PLC), or a broken power-law EED is commonly
adopted \citep[e.g.,][]{tave98,finke08}. The main justification
for this EED approximation is that the non-thermal emissions from
blazars can be described by a power-law spectrum, and the
power-law EED can be naturally generated in the
framework of the shock acceleration (the Fermi I process)
\citep[e.g.,][]{baring97}. In recent observations, however, the
X-ray spectra of several blazars (like Mrk 421, Mrk 501) show
significant curvature, which are typically milder than an
exponential cut-off \citep{Massaro1,Massaro2,Massaro3}. Very
recently, it has been found that gamma-ray emissions of many
blazars are successfully fitted with a log-parabolic (LP) spectrum
\citep[e.g.,][]{Aharonian09,Abdo10,Ackermann11}. The LP EED is
then proposed to model the observed spectral curvature, and such
LP EED can be generated in the stochastic turbulence acceleration
scenario (the Fermi II process) \citep[e.g.,][]{Tramacere09,Tramacere11}.
Numerical simulation indicates that stochastic acceleration
process may play an important role in the formation of the
particle spectrum in blazar jet \citep{V05}. When the emission
mechanisms are determined, the emitting EED can be reconstructed
from the observed emission spectra. We can then investigate the
acceleration processes acting in blazar jet
\citep[e.g.,][]{Ushio,garson}.

Blazars are well known for their rapid and large-amplitude variability at all wavebands,
most prominently at keV and TeV energies. The relations (including correlation and time lag)
between variabilities at different wavelengths are crucial for constraining jet models
\citep[e.g.,][]{Sokolov,Fossati08,Kata10,bott10}.
For instance, if the hard lag is
observed, a acceleration process could be considered in the model.
Different flare patterns indicate different causes of the flare, such as change in the injection
rate and change in the acceleration process, etc \citep[e.g.,][]{kirk,kataoka,Graff,Maraschi,chen}.

Mrk 421 is the closest known (redshift $z=$0.031) and the first
very high energy (VHE) blazar \citep{Punch}. It is classified as
high-peaked BL Lac (HBL) according to its synchrotron peak
location. Mrk 421 is the main target of multi-wavelength
monitoring campaigns. There are a large number of publications on
the multi-wavelength observations of this source. Its SED and
relation of variabilities at different bands are intensively
studied \citep[e.g.,][]{bla,alb,acc,acc11,ale,argo}. From these
studies, we learned that TeV flare is often correlated with keV
flare, however, such relation is very complex. For example, the
TeV flux increases more than quadratically with respect to the X-ray one, and
there is time lag between TeV and keV flares. Although SED of Mrk
421 is commonly well fitted by a one-zone SSC model, the complex
variability behaviors remind us that the realistic model seems
more complicated. Lately, the multi-wavelength campaign showed
evidence for Mrk 421 in low/quiescent state from 2009 January 19
to 2009 June 1 \citep{Abdo11}. During this campaign, no
significant flare activity was seen and the measured VHE flux is
among the lowest fluxes recorded by MAGIC. \citet{Abdo11}
therefore claimed that the unprecedented complete, 4.5 months
average SED observed during this campaign can be considered as an
excellent proxy for the low/quiescent state SED of Mrk 421.
Several months later, Mrk 421 was found to undergo its one of the
brightest outburst at X-rays and gamma-rays bands on February 17,
2010 \citep{Shukla}. During this flaring, the flux correlation at
X-ray and TeV band was observed \citep{Shukla}, and intra-night
variability at GeV - TeV band was found
\citep{Galante,Shukla,Raue}.

In this paper we investigate the EEDs and the acceleration
processes in the jet of Mrk 421 in low state and giant flare
state. To achieve our aim, we assume two electron distributions,
both well motivated by the current particle acceleration models:
the shock accelerated PLC EED and the stochastic turbulence
accelerated LP EED, to model the SEDs in the frame of a one-zone
SSC model. To more efficiently constrain the model parameters and
better distinguish between the models, we employ the Markov Chain
Monte Carlo (MCMC) method instead of a simple
$\chi^2$-minimization procedure to investigate the
high-dimensional model parameter space systematically. The
emission models and MCMC method are briefly described in Section
\ref{model}. In Section \ref{result}, we report our results.
Finally in Section \ref{DC} we summarize our discussions and
conclusions.

\section{Emission model and MCMC method}
\label{model}

\subsection{Emission model}

The one-zone SSC assumes that non-thermal radiation is produced by
both the synchrotron radiation and SSC process in a spherical blob
filled with uniform magnetic field ($B$), which is moving
relativistically at a small angle to our line of sight, and the
observed radiation is strongly boosted by a relativistic Doppler
factor $\delta_{\rm D}$. The radius of emitting blob is
$R^{\prime}=\frac{t_{\rm v,min}\delta_Dc}{1+z}$, where $t_{\rm
v,min}$ is the minimum variability timescale. Here, quantities in
the observer's frame are unprimed, and quantities in the comoving
frame are primed. Note that the magnetic field $B$ is defined in
the comoving frame, despite being unprimed. We use the methods
given in \citet{finke08} to calculate synchrotron and SSC fluxes.

The PLC electron distribution formed in the Fermi I acceleration process is
\begin{equation}
N^{\prime}(\gamma^{\prime})=K^{\prime}_{\rm e}\gamma^{\prime
-s}\exp(\frac{-\gamma^{\prime}}{\gamma^{\prime}_c})\ \ {\rm for}\
\ \gamma^{\prime}_{\rm
min}\leq\gamma^{\prime}\leq\gamma^{\prime}_{\rm max},
\end{equation}
where $K^{\prime}_{\rm e}$ is the normalization of the EED, $s$ is
the electron energy index, and $\gamma^{\prime}_c$ is the electron
cut-off energy. $\gamma^{\prime}_{\rm min}$ and
$\gamma^{\prime}_{\rm max}$ is the electron minimum energy and
electron maximum energy, respectively. In this model, there are
eight free parameters, five of them  specify the electron energy
distribution ($K^{\prime}_e$, $\gamma^{\prime}_{\rm min}$,
$\gamma^{\prime}_{c}$, $\gamma^{\prime}_{\rm max}$, $s$) and the other
three ones describe the global properties of the emitting region ($B$,
$R^{\prime}$, $\delta_D$).

The LP electron distribution generated in the Fermi II acceleration process is
\begin{equation}
 N^{\prime}(\gamma^{\prime})= K^{\prime}_e~\left\{
 \begin{array}{ll}
\left(\frac{\gamma^{\prime}}{\gamma^{\prime}_c}\right)^{-s} & \gamma^{\prime}_{\rm min}\leq \gamma^{\prime}\leq\gamma^{\prime}_{c} \\
\left(\frac{\gamma^{\prime}}{\gamma^{\prime}_c}\right)
 ^{-[s+r\log(\frac{\gamma^{\prime}}{\gamma^{\prime}_c})]} &  \gamma^{\prime}_c\leq\gamma^{\prime}\leq\gamma^{\prime}_{\rm
 max}\;,
 \end{array}
 \right.
\end{equation}
where $r$ is the curvature term of EED \citep{Massaro3}. In this
model, $r$ is an additional one besides the eight parameters
mentioned above.

\subsection{MCMC method}

The MCMC method, based on the Bayesian statistics, is well
suitable for high dimensional parameter space investigation, which
is superior to the grid approach with a more efficient sampling of
the parameter space of interest. The Metropolis-Hastings sampling
algorithm is adopted to determine the jump probability from one
point to the next in the parameter space \citep{Mackay}. The
algorithm ensures that the probability density functions (PDF) of
model parameters can be asymptotically approached with the number
density of samples. In the following fitting, we will run single
chains using the Raftery \& Lewis convergence diagnostics
\citep{RL}, and we assume flat priors in the model parameters
spaces. A brief introduction to the basic procedure of the MCMC
sampling can be found in \citet{Fan}. For more details about the
MCMC method, please refer to \citet{Neal1993,Gamerman1997,
Mackay}. Since the code we used in this paper \citep{Liuj} is
adapted from COSMOMC, we refer the reader to the
website\footnote{\texttt{http://cosmologist.info/cosmomc/}} and to
\citet{Lewis} and references therein for a detailed explanation of
the code about the sampling options, convergence criteria and
statistical quantities.

\section{Modeling results}
\label{result}

\subsection{Modeling the SED in low state}
\label{lowsed}

In this case, we adopt the optical-UV to X-ray data observed by
{\it Swift}/UVOT/RXT/BAT and the {\it Fermi}-LAT gamma-rays data
reported in \citet{Abdo11}. For the {\it Fermi}-LAT data, the last
two data points are not included in our modeling, due to their
very large uncertainties. Instead, we use the first two data
points measured by MAGIC since the extragalactic background light
(EBL) absorption on the flux at such energy band is negligible.
Because the EBL absorption has an effect on the other fluxes
measured by MAGIC, we also do not take them into account in our
modeling. The absorption effect of EBL will be discussed in \S
3.3. The SMA data at $2.3\times10^{11}$ Hz are used to constrain
$\gamma^{\prime}_{\rm min}$. We fix $\gamma^{\prime}_{\rm
min}=700$ for PLC case and $\gamma^{\prime}_{\rm min}=1200$ for LP
case and $\gamma^{\prime}_{\rm max}=10^8$ for both case.

\begin{figure*}[]
  \begin{center}
  \begin{tabular}{ccc}
\hspace{-0.90cm}
     \includegraphics[width=60mm,height=60mm]{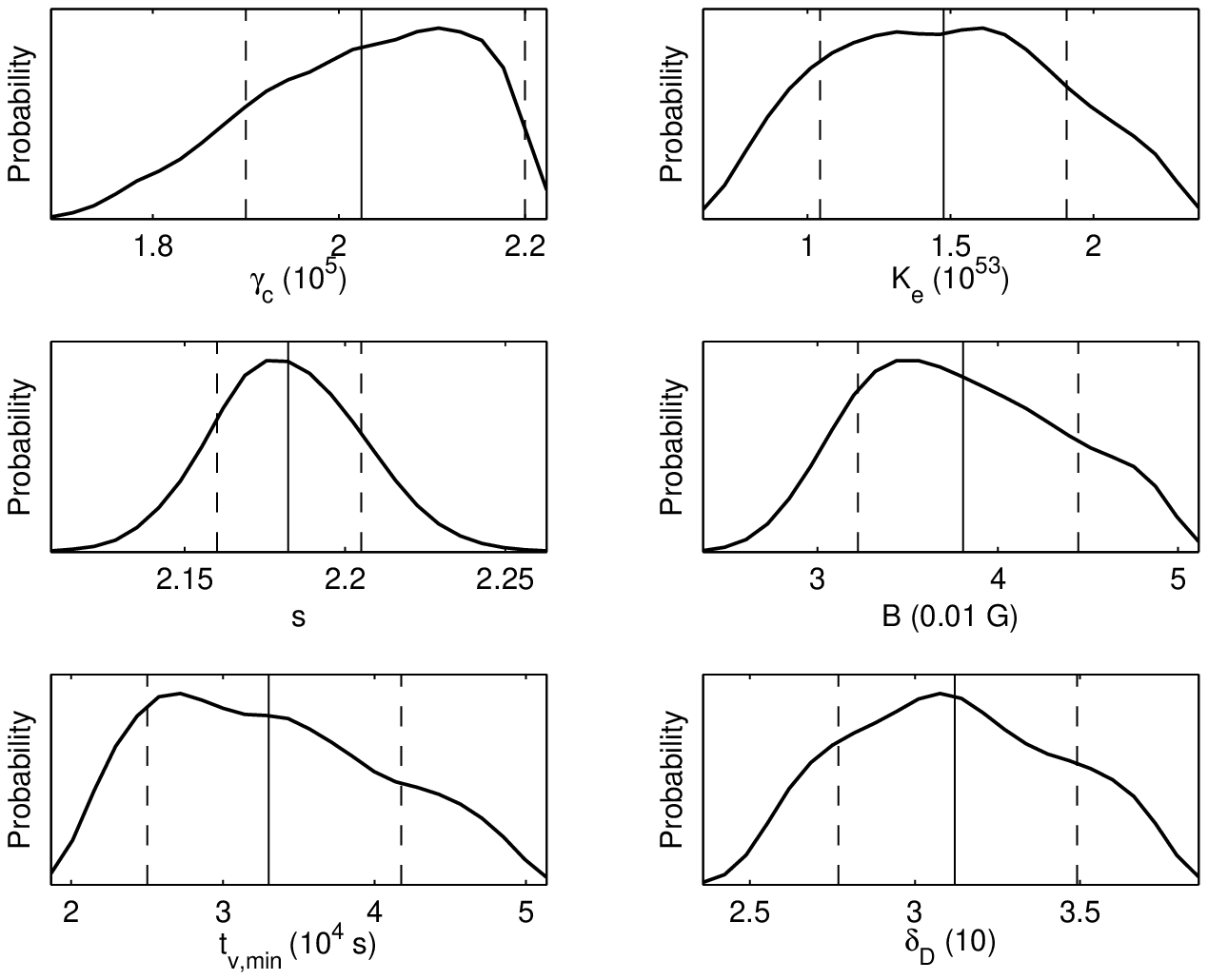} &
\hspace{-0.90cm}
     \includegraphics[width=60mm,height=60mm]{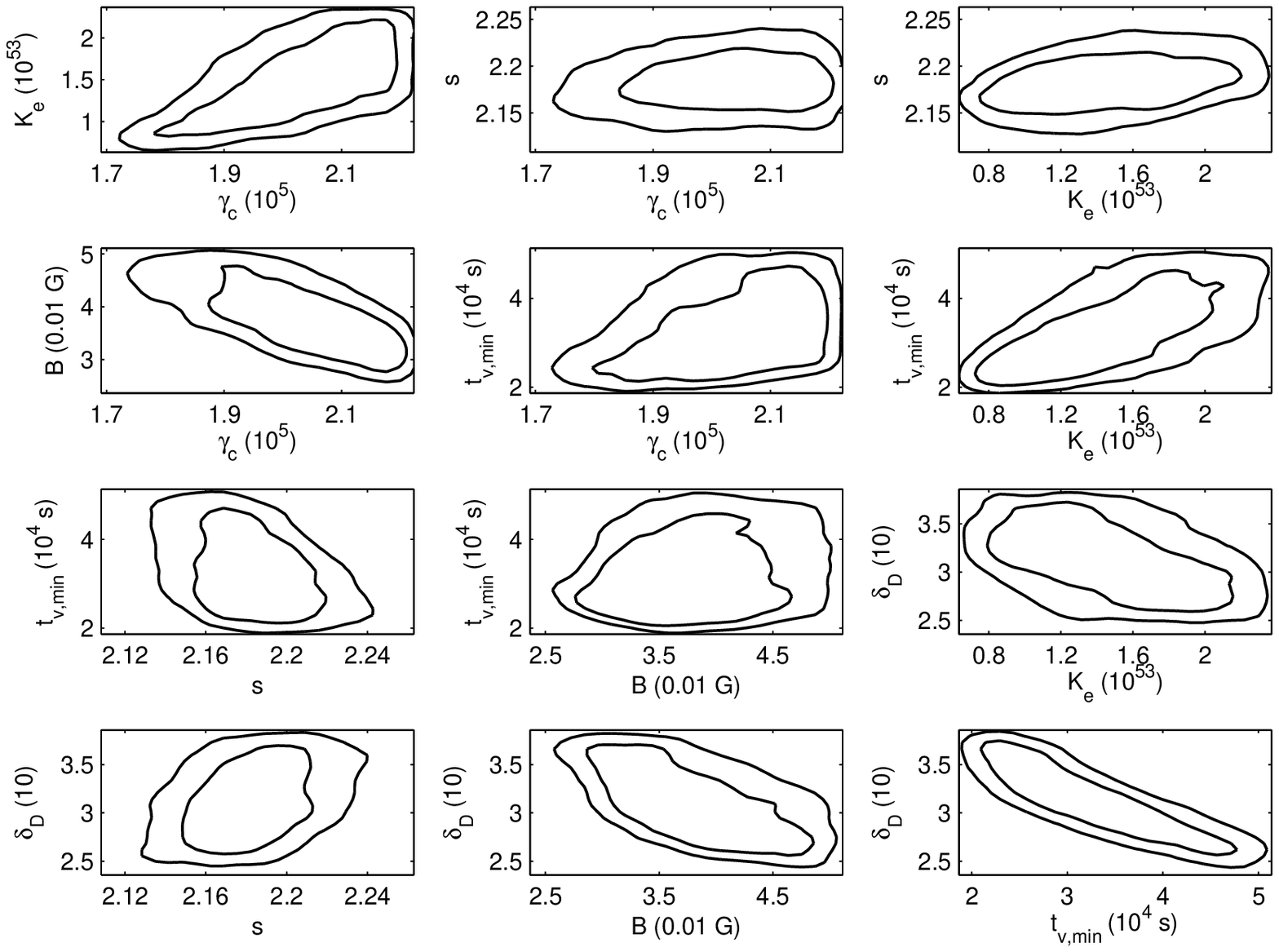} &
\hspace{-0.90cm}
     \includegraphics[width=60mm,height=60mm]{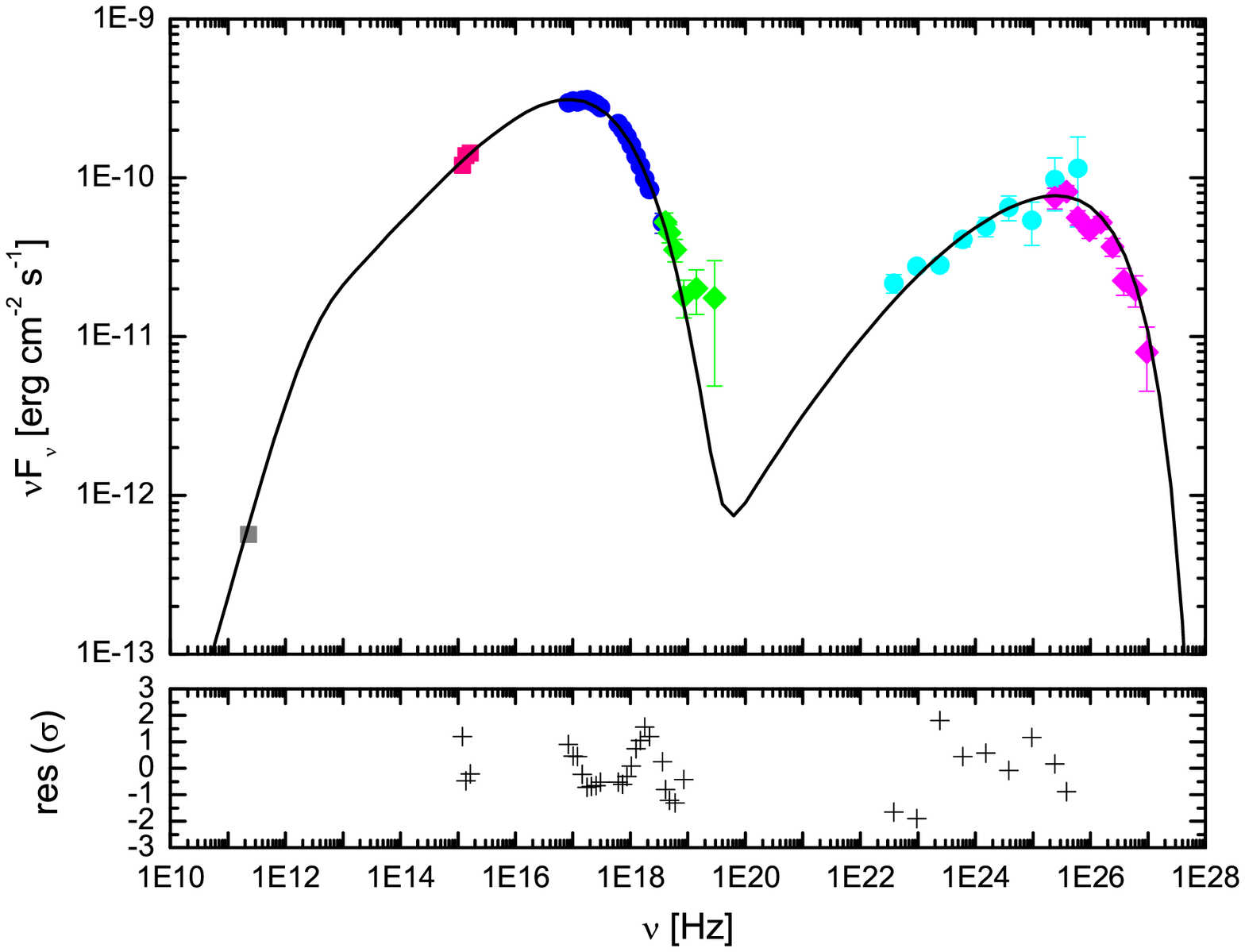}
\end{tabular}
  \end{center}
\caption{Modeling the SED in low state with the PLC electron
distribution. Left: 1-D marginalized probability distribution of
the parameters. The solid vertical line is the expected value, and
the 68\% limits are depicted by the dashed vertical lines; Middle:
2-D confidence contours of the parameters. The contours are for 1
and 2$\sigma$ levels; Right: the best fit to the SED from
optical-UV-X-ray to GeV bands. For the low energy component, the
symbols represent {\it Swift}/UVOT (squares), RXT (circles) and
BAT (diamonds) data; For high energy component, the symbols
represent {\it Fermi}-LAT (circles) and MAGIC (diamonds) data. For
the detailed information of the observed data sets, please refer
to \citet{Abdo11}. } \label{fig:lowEc}
\end{figure*}

For the PLC electron distribution, the one-dimensional (1D)
probability distributions and two-dimensional (2D) confidence
regions (at 1$\sigma$ and 2$\sigma$ confidence levels) of the
model parameters, and the SED are shown in Fig.~\ref{fig:lowEc}.
The fitting parameters are summarized in Table~\ref{table:low}
with the reduced $\chi^2_{\nu}=1.01$ for 27 d.o.f. (Note that a
relative systematic uncertainty of 5\% was added in quadrature to
the statistical error of the optical-UV-X-rays data reported in
\citet{Abdo11}). It can be seen that a very good constraint is
derived for the spectral energy index $s$. The parameters
$\gamma^{\prime}_{c}$, $K^{\prime}_e$ and $\delta_{\rm D}$ are
well constrained. The constraints on $B$ and $t_{\rm v,min}$ are
not strong, but they are still restricted in relatively small
ranges.

For the 2D confidence regions of the parameters, we only show some
combinations with relatively large correlations. It can be seen
that there are  negative correlations between $\gamma'_c$ and $B$,
$B$ and $\delta_{\rm D}$, as well as $t_{\rm v,min}$ and $\delta_{\rm
D}$, but there are positive correlations between $K'_e$ and
$t_{\rm v,min}$ and between $K'_e$ and $\gamma'_c$. In the
following, we try to physically understand the correlations among
some parameters. In the standard SSC model, the synchrotron SED
peak flux in the $\delta$-approximation is $\nu F_{\nu}\propto
N^{\prime}(\gamma^{\prime}_{\rm c})B^2\gamma^{\prime 3}_{\rm c}\delta^4_{\rm D}$
and the synchrotron peak energy is $E_{\rm s}\propto\gamma^{\prime
2}_{\rm c} B \delta_{\rm D}$ \citep[e.g.,][]{Tramacere11}. The
negative correlation between $\gamma^{\prime}_{\rm c}$ and $B$ is
predicted for the constant $E_{\rm s}$ which can be estimated by
the observed spectra. Because $B\delta_{\rm D}\propto \nu^2_{\rm
s}/\nu_{\rm c}$ in the SSC model in the Thomson regime \citep[e.g.,][]{tave98}, where
$\nu_{\rm s}$ and $\nu_{\rm c}$ are the peak frequencies of the
synchrotron and IC scattering respectively, the negative
correlation between $B$ and $\delta_{\rm D}$ is expected for the
observed $\nu_{\rm s}$ and $\nu_{\rm c}$, which implies that
the IC scattering is in the Thomson regime. In the SSC model, the
radius $R^{\prime}$ of the emitting blob relates to the gamma-ray flux
($\propto R^{\prime-2}$) and then can be estimated using the observed
gamma-ray flux, meanwhile $R^{\prime}\sim c\delta_{\rm D}t_{\rm
v,min}$, hence a negative correlation between $t_{\rm v,min}$ and
$\delta_{\rm D}$ can be predicted. On the other hand, with
$B\propto E_{\rm s}/(\gamma^{\prime 2}_{\rm c}\delta_{\rm D})$,
the synchrotron peak flux can be rewritten as $(\nu F_{\nu})_{\rm
s}\propto (K^{\prime}_{\rm e}/\gamma^{\prime}_{\rm
c})\delta^2_{\rm D}E^2_{\rm s}$, therefore the positive correlation
between $K^{\prime}_{\rm e}$ and $\gamma^{\prime}_{\rm c}$ is
expected by the constraint of the observed synchrotron peak flux.
As to the positive correlation between $K^{\prime}_e$ and $t_{\rm
v,min}$, since IC peak flux $(\nu F_\nu)_{\rm c}$ can be estimated
by using the observed gamma-ray flux and $(\nu F_\nu)_{\rm c}/(\nu
F_{\nu})_{\rm s}\propto K'_e/\delta^2_{\rm D}t^2_{\rm v,min}$,
and then $K'_e/t_{\rm v,min}\propto [(\nu F_\nu)_{\rm c}/(\nu F_{\nu})_{\rm
s}][\nu_{\rm c}/\nu^2_{\rm s}]$ where $t_{\rm
v,min}B\delta^3_{\rm D}=\rm constant$ and $B\delta_{\rm D}\propto
\nu^2_{\rm s}/\nu_{\rm c}$ \citep{tave98} are used. Therefore, the
positive correlation between $K^{\prime}_e$ and $t_{\rm v,min}$
can be predicted according to the observations.

\begin{figure*}[]
  \begin{center}
  \begin{tabular}{ccc}
\hspace{-0.89cm}
     \includegraphics[width=60mm,height=60mm]{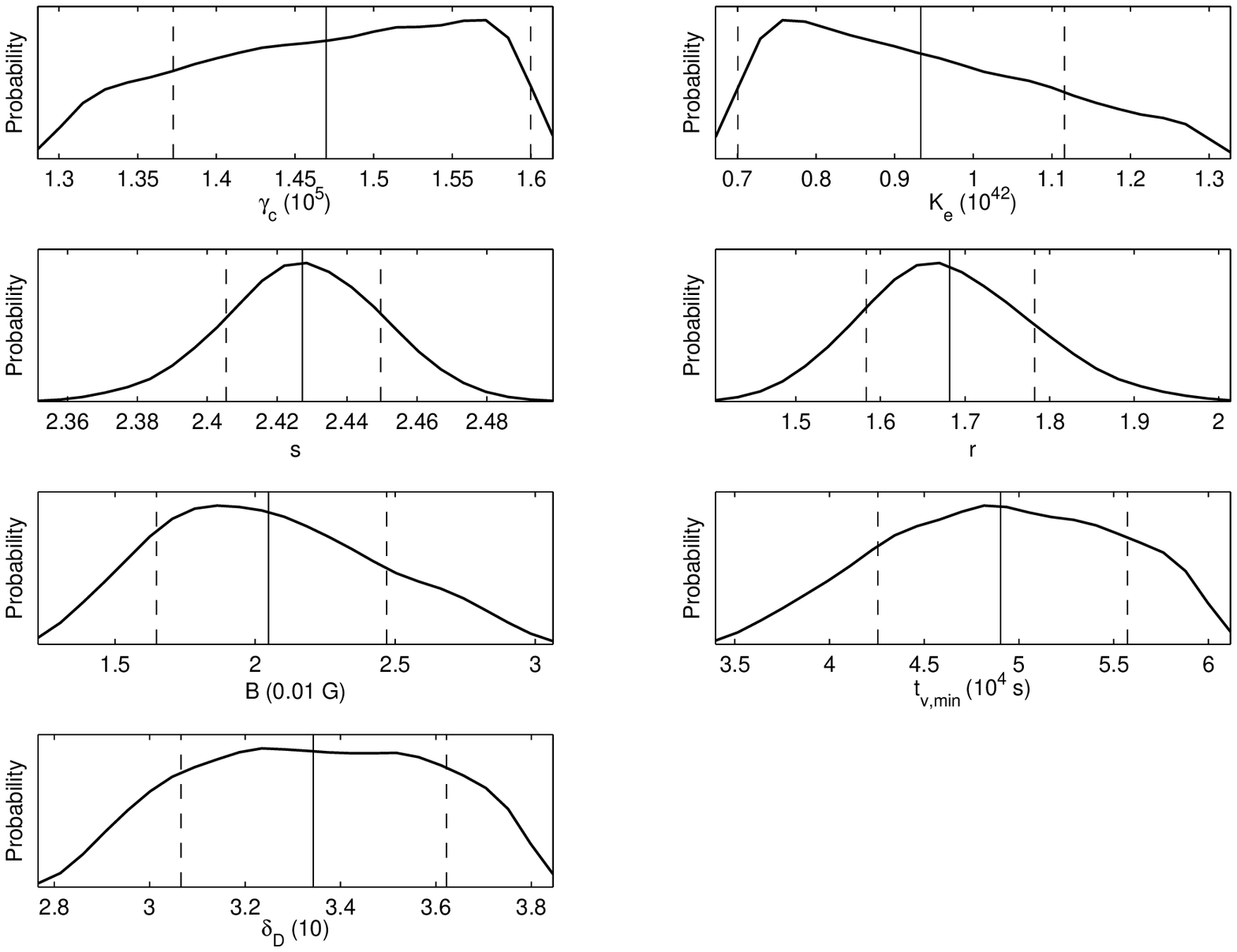} &
\hspace{-0.89cm}
     \includegraphics[width=60mm,height=60mm]{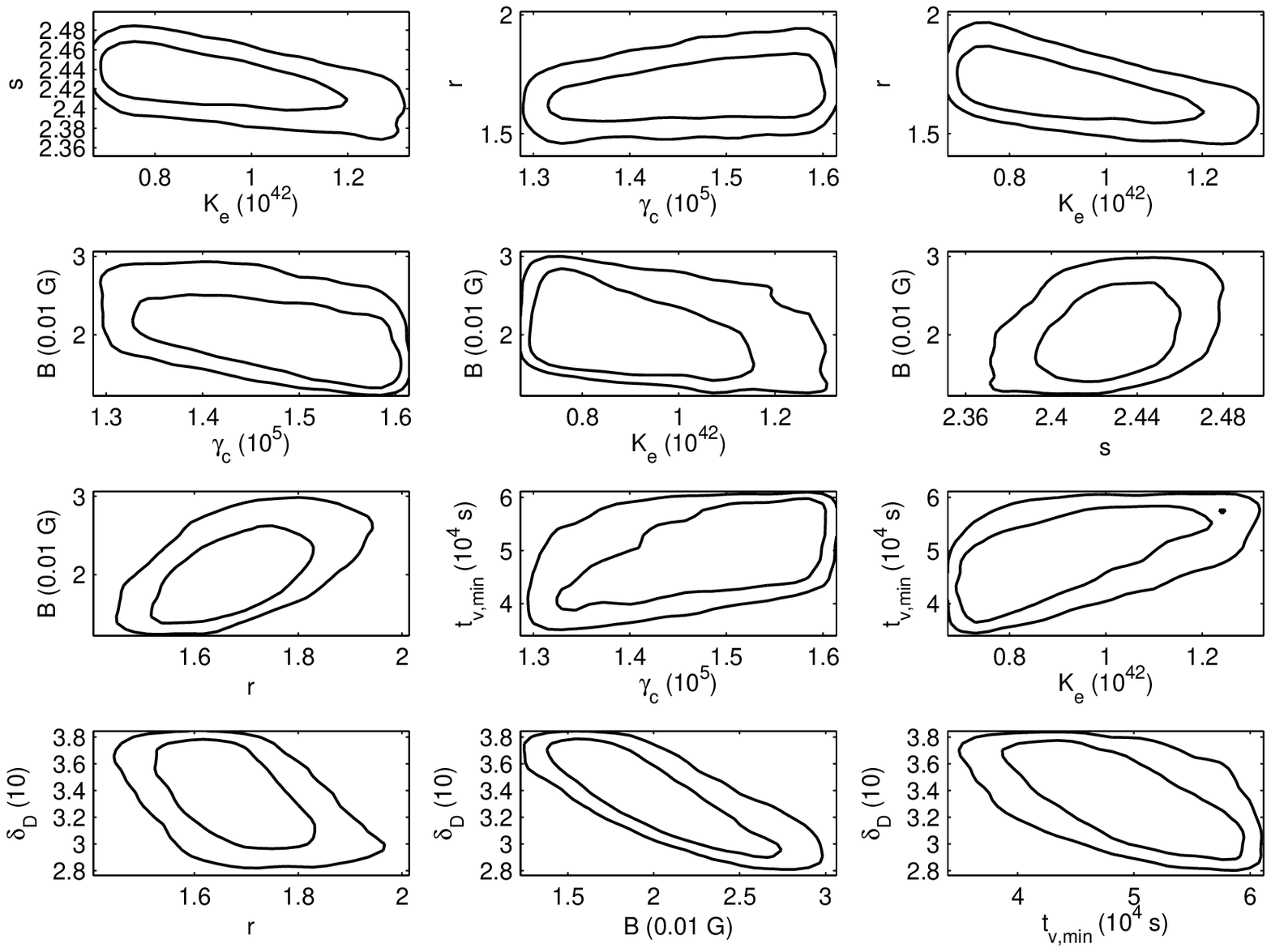} &
\hspace{-0.89cm}
     \includegraphics[width=60mm,height=60mm]{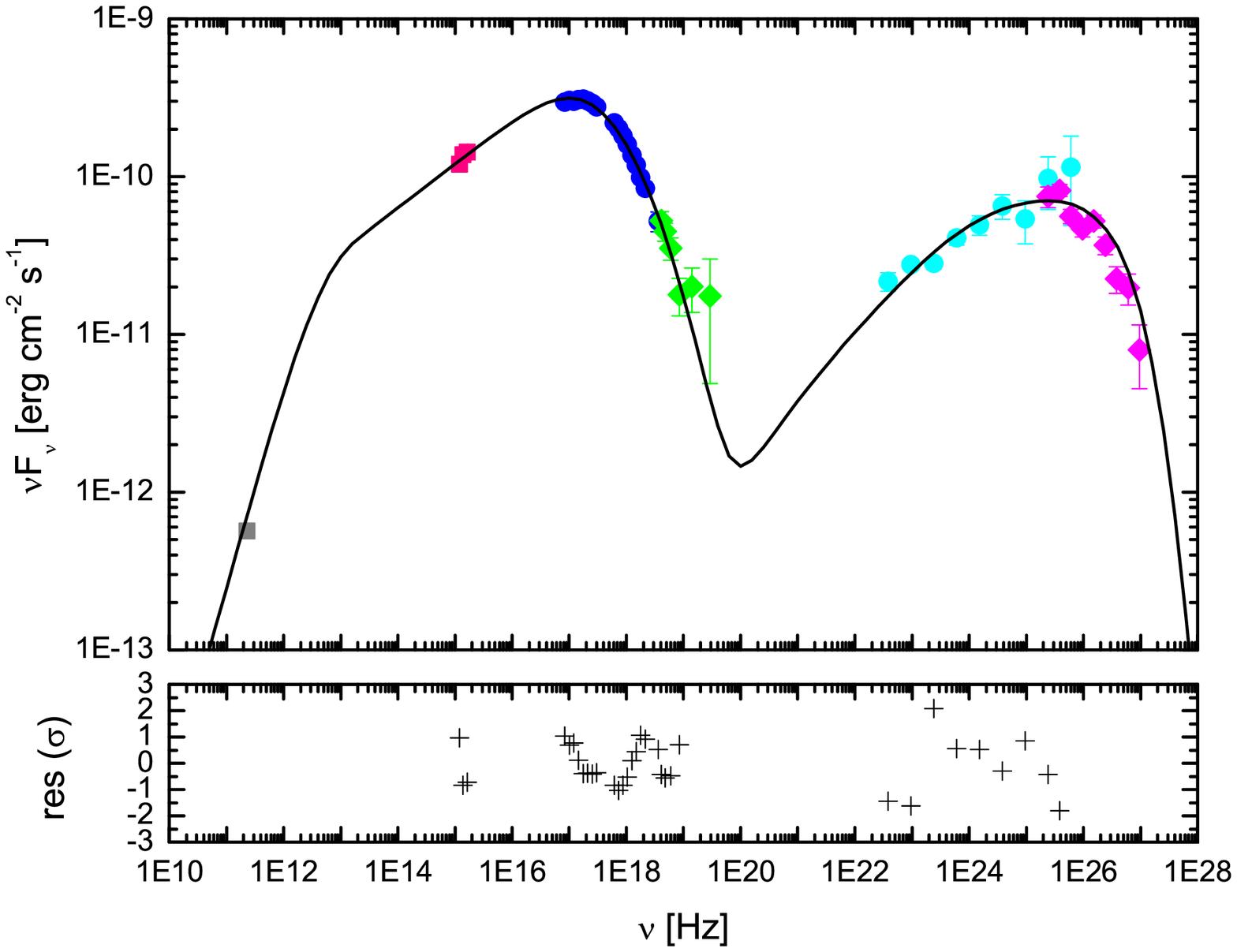}
\end{tabular}
\end{center}
\caption{Same as in Fig.~\ref{fig:lowEc} but with the LP electron distribution.
}
\label{fig:lowlp}
\end{figure*}

For the LP electron distribution, the parameters distributions and
SED are shown in Fig.~\ref{fig:lowlp}. The fitting parameters are
listed in Table~\ref{table:low}. In this case, we have
$\chi^2_{\nu}=0.97$ (for 26 d.o.f.). The curvature term $r$ is
well constrained. It seems that $B$ can be better constrained than
that in PLC model. The constraints on the other parameters are
similar to that derived in the PLC model. The explanations on the
correlations in the case of PLC still hold in the case of LP
except for the correlation between $B$ and $\gamma^{\prime}_{c}$.
In LP case, the correlation between $B$ and $\gamma^{\prime}_{c}$
becomes weak, which is caused by the effect of $r$ . This
correlation between $B$ and $r$ is caused by the well-observed
spectrum around the synchrotron peak.

Comparing the results derived in the two EEDs, it can be found
that the global properties of the emitting blob ($B$, $\delta_{\rm
D}$, $R^{\prime}$) derived in the two cases are comparable.
Obviously, we cannot distinguish between the PLC and LP electron
distributions in the low state based on the above results (see
Figs.~\ref{fig:lowEc} and \ref{fig:lowlp} as well as
Table~\ref{table:low}).

\begin{deluxetable}{ccccccccc}
\centering
\tabletypesize{\scriptsize}
\tablecaption{Fitting parameters for the SED in low state.}
\tablewidth{0pt}
\tablehead{
\colhead{Model} & \colhead{$\gamma^{\prime}_{c}$} & \colhead{$K^{\prime}_{e}$} &
\colhead{$s$} & \colhead{$r$} & \colhead{$B$} & \colhead{$t_{\rm v,min}$} & \colhead{$\delta_{\rm D}$} & \colhead{$U^{\prime}_e/U^{\prime}_B$\tablenotemark{b}} \\
 & \colhead{($10^5$)} & \colhead{($10^{53}$/$10^{42}$)\tablenotemark{a}}  & & & \colhead{(0.01 G)} &
 \colhead{($10^4$ s)} & \colhead{(10)} &
 }
\startdata
PLC model & 2.02$\pm$0.11  & 1.48$\pm$0.39  & 2.19$\pm$0.02  & -- & 3.81$\pm$0.56  & 3.30$\pm$0.75  & 3.12$\pm$0.32 & 34.61 \\
  68\% limit & (1.90 - 2.20) & (1.04 - 1.91) & (2.16 - 2.21) & -- & (3.22 - 4.45) & (2.50 - 4.18) & (2.77 - 3.49) \\
  \tableline
LP model & 1.47$\pm$0.08  & 0.93$\pm$0.16  & 2.43$\pm$0.02  & 1.68$\pm$0.10 & 2.05$\pm$0.38  & 4.90$\pm$0.60  & 3.34$\pm$0.25 & 50.71 \\
  68\% limit & (1.37 - 1.60) & (0.70 - 1.12) & (2.41 - 2.45) & (1.58 - 1.78) & (1.65 - 2.47) & (4.26 - 5.57) & (3.07 - 3.62) \\
\enddata
\tablenotetext{a}{$10^{53}$ for PLC and $10^{42}$ for LP.}
\tablenotetext{b}{Ratio of energy density in electrons to that in magnetic field.}
\label{table:low}
\end{deluxetable}

\subsection{Modeling the SED in flare state}

\begin{figure*}[]
  \begin{center}
  \begin{tabular}{ccc}
\hspace{-0.89cm}
     \includegraphics[width=60mm,height=60mm]{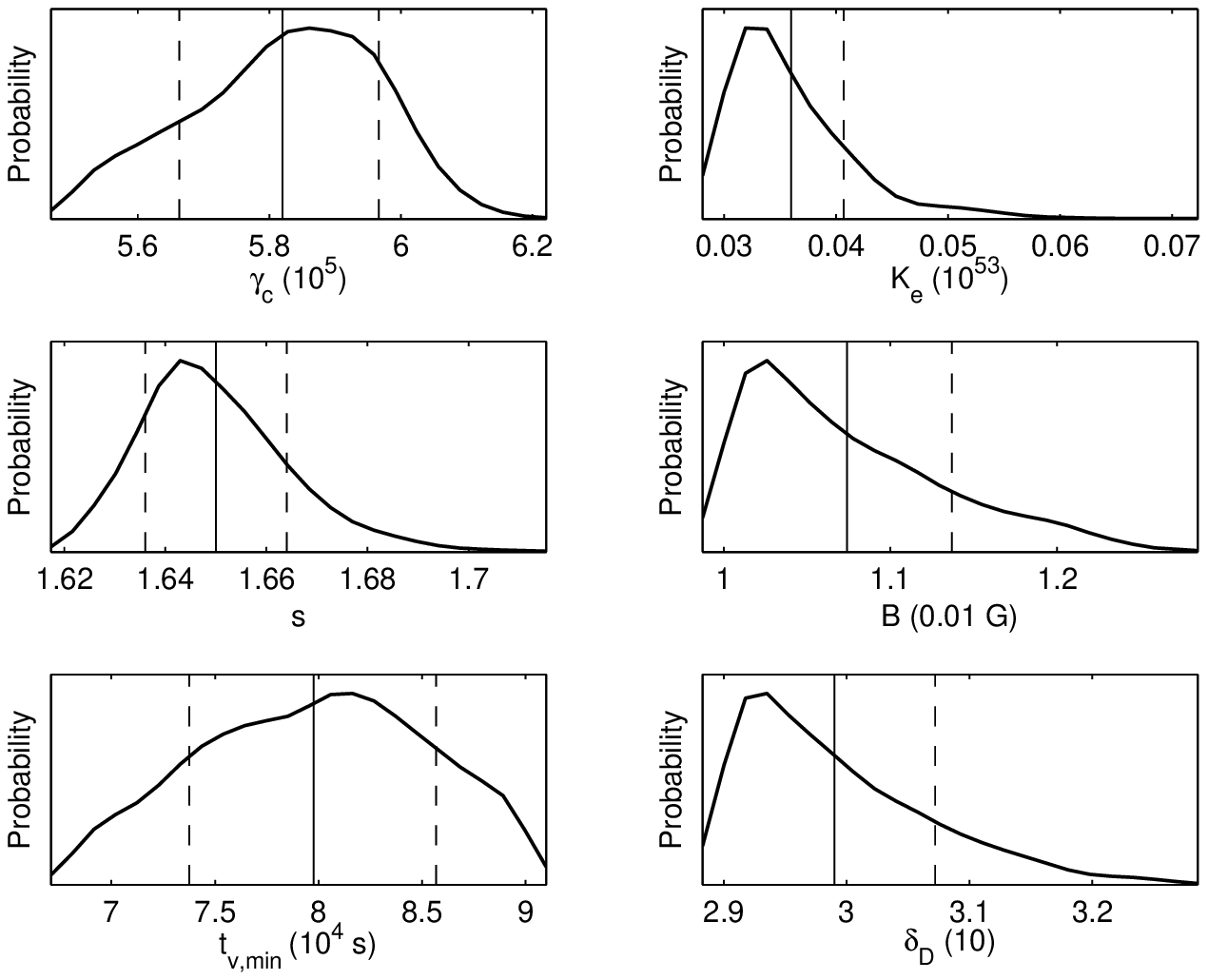} &
\hspace{-0.89cm}
     \includegraphics[width=60mm,height=60mm]{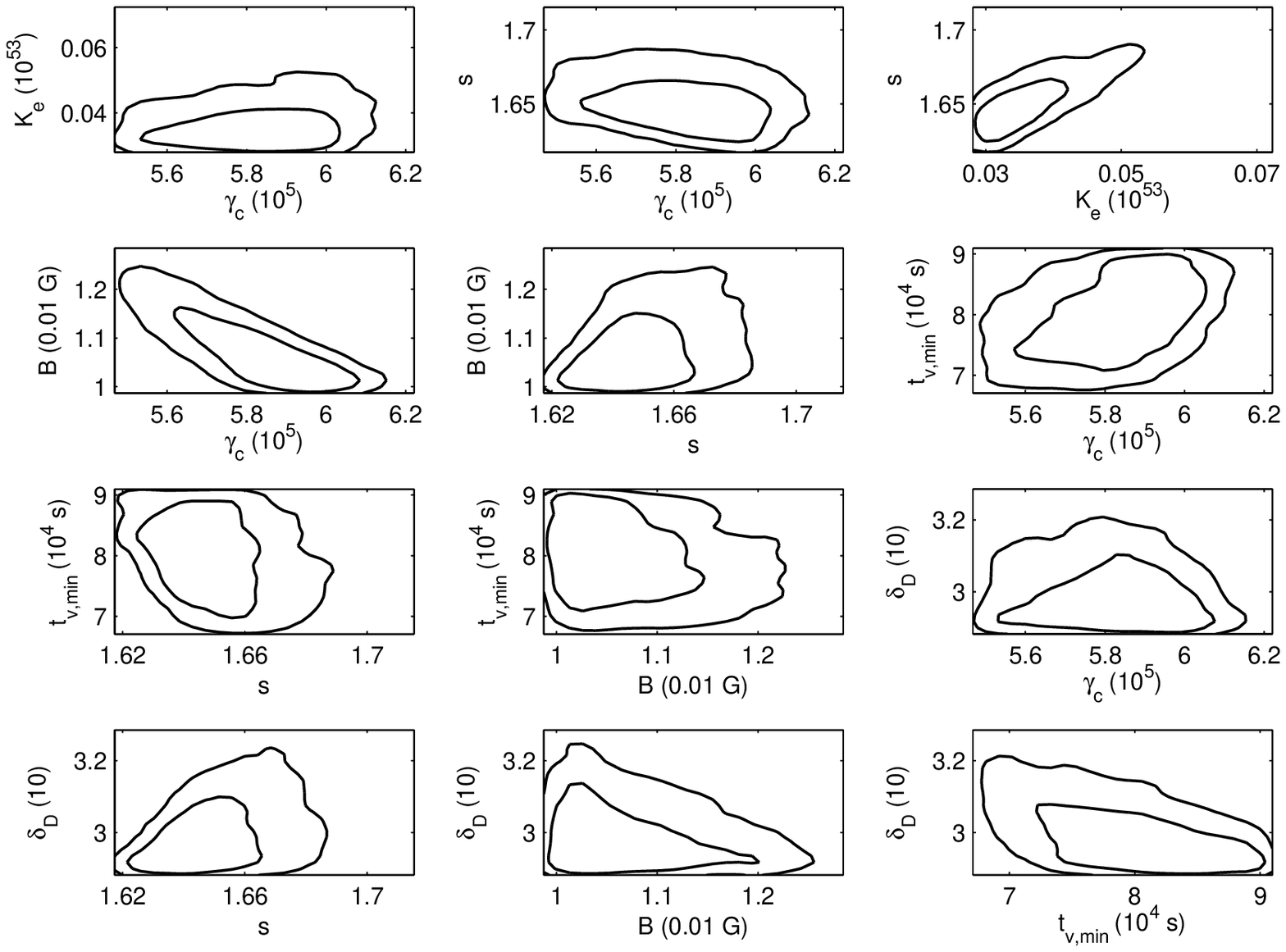} &
\hspace{-0.89cm}
     \includegraphics[width=60mm,height=60mm]{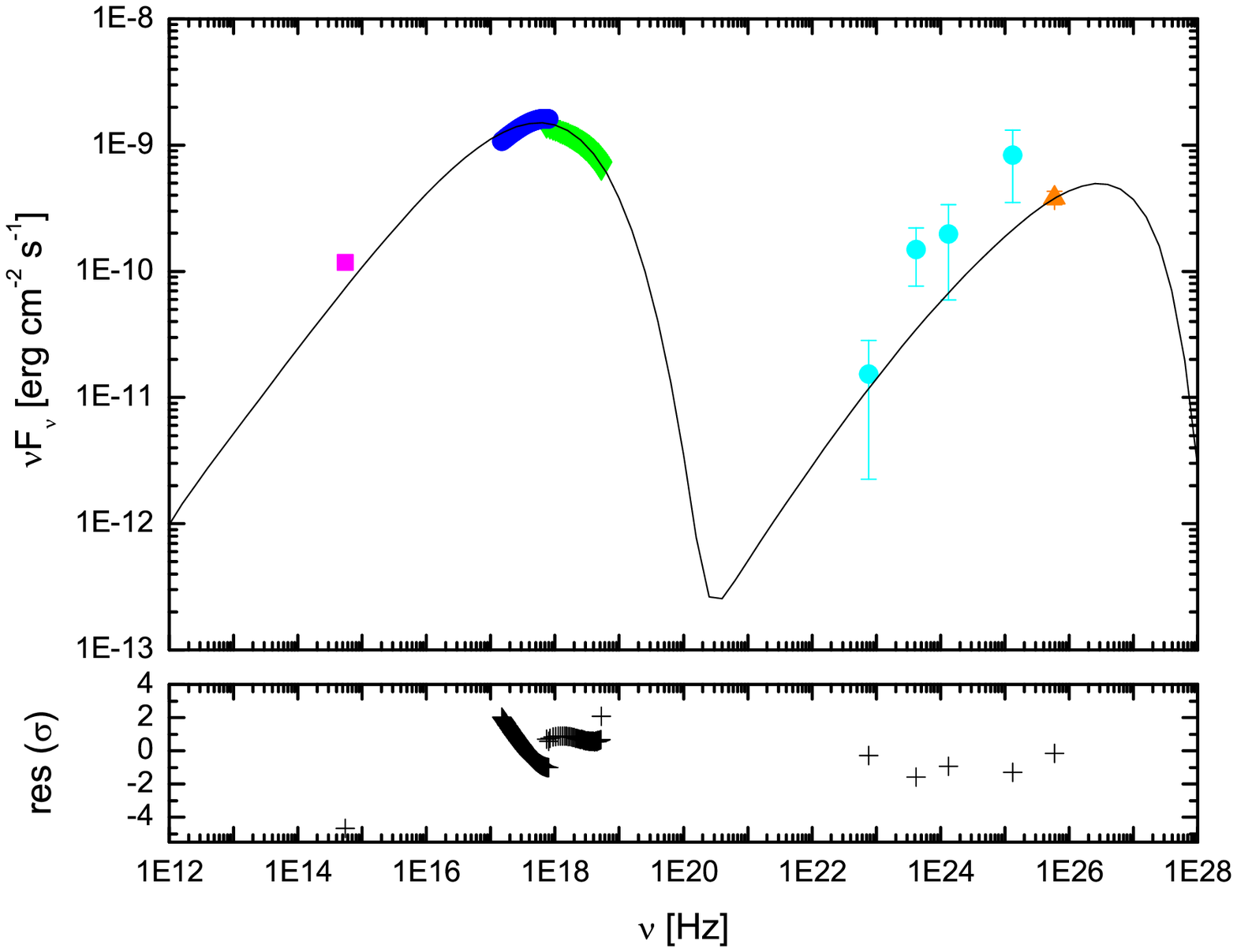}
\end{tabular}
  \end{center}
\caption{Modeling the SED in giant flare state with the PLC
electron distribution. The curves are the same as that in
Fig.~\ref{fig:lowEc}. For low energy component of SED, the symbols
denote the data from SPOL CCD Imaging/Spectropolarimeter at
Steward Observatory (squares), the XRT data (circles) and the PCA
data (triangles); At high energy component, the symbols denote LAT
data (circles) and HAGAR data (triangle). Please see
\citet{Shukla} for more details about the data sets. }
\label{fig:highEc}
\end{figure*}

For the SED in flaring state, the {\it Swift}/RXT, {\it RXTE}/PCA,
{\it Fermi}-LAT and HAGAR data reported in \citet{Shukla} are
adopted. We also fix $\gamma^{\prime}_{\rm min}$ and
$\gamma^{\prime}_{\rm max}$ as we did in Section~\ref{lowsed}. The
parameters distributions and SED fitting derived with PLC EED are
shown in Fig.~\ref{fig:highEc}. The fitting parameters are listed
in Table~\ref{table:high}. The resulting $\chi^2_{\nu}=0.79$ for
318 d.o.f. is below unity after a relative systematic uncertainty
of 8\% was added. The fittings at optical and GeV bands are bad.
$\gamma^{\prime}_{c}$ and $s$ are well constrained, and $t_{\rm
v,min}$ can be constrained in relatively small range. However, the
parameters $K^{\prime}_{e}$, $\delta_{\rm D}$ and $B$ are poorly
constrained. We cannot obtain the meaningful distribution ranges
of the three parameters, thus only the 68\% upper limit are
reported in Table~\ref{table:high}. Unfortunately, the variability
timescale ($\sim1$ day) required in this PLC model contradicts the
observed intra-day variability at GeV - TeV bands
\citep{Shukla,Raue}.

\begin{figure*}[]
  \begin{center}
  \begin{tabular}{ccc}
\hspace{-0.89cm}
     \includegraphics[width=60mm,height=60mm]{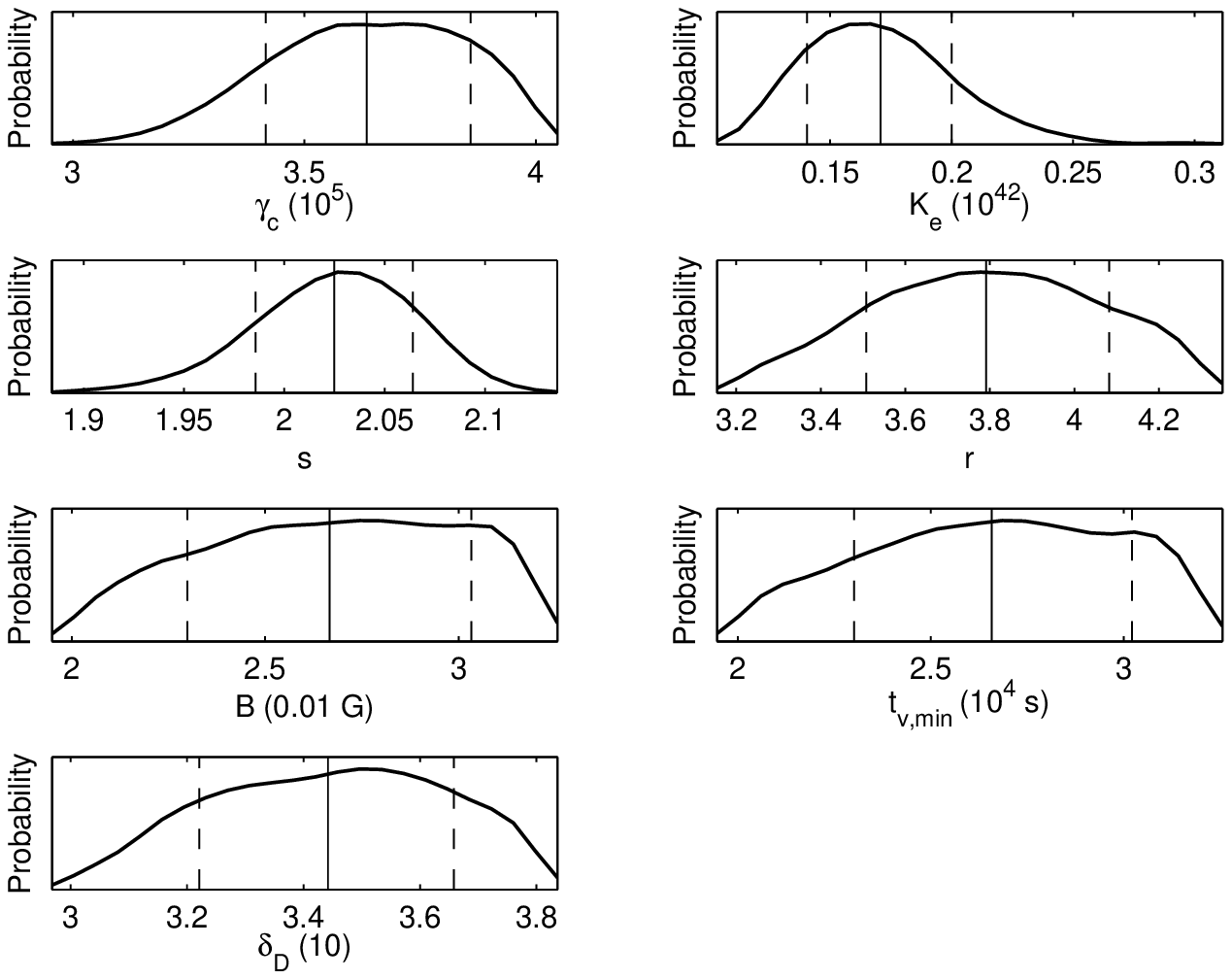} &
\hspace{-0.89cm}
     \includegraphics[width=60mm,height=60mm]{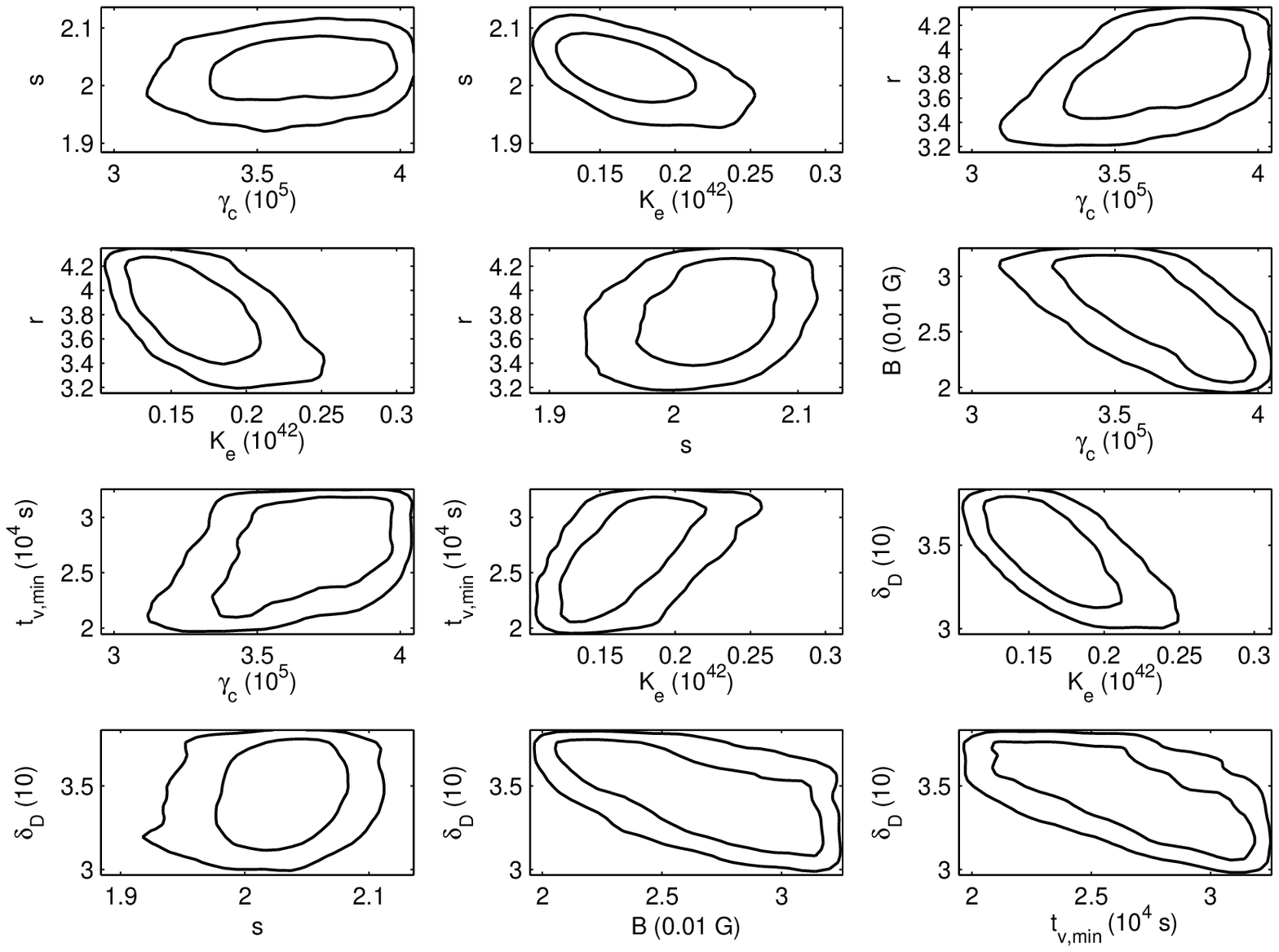} &
\hspace{-0.89cm}
     \includegraphics[width=60mm,height=60mm]{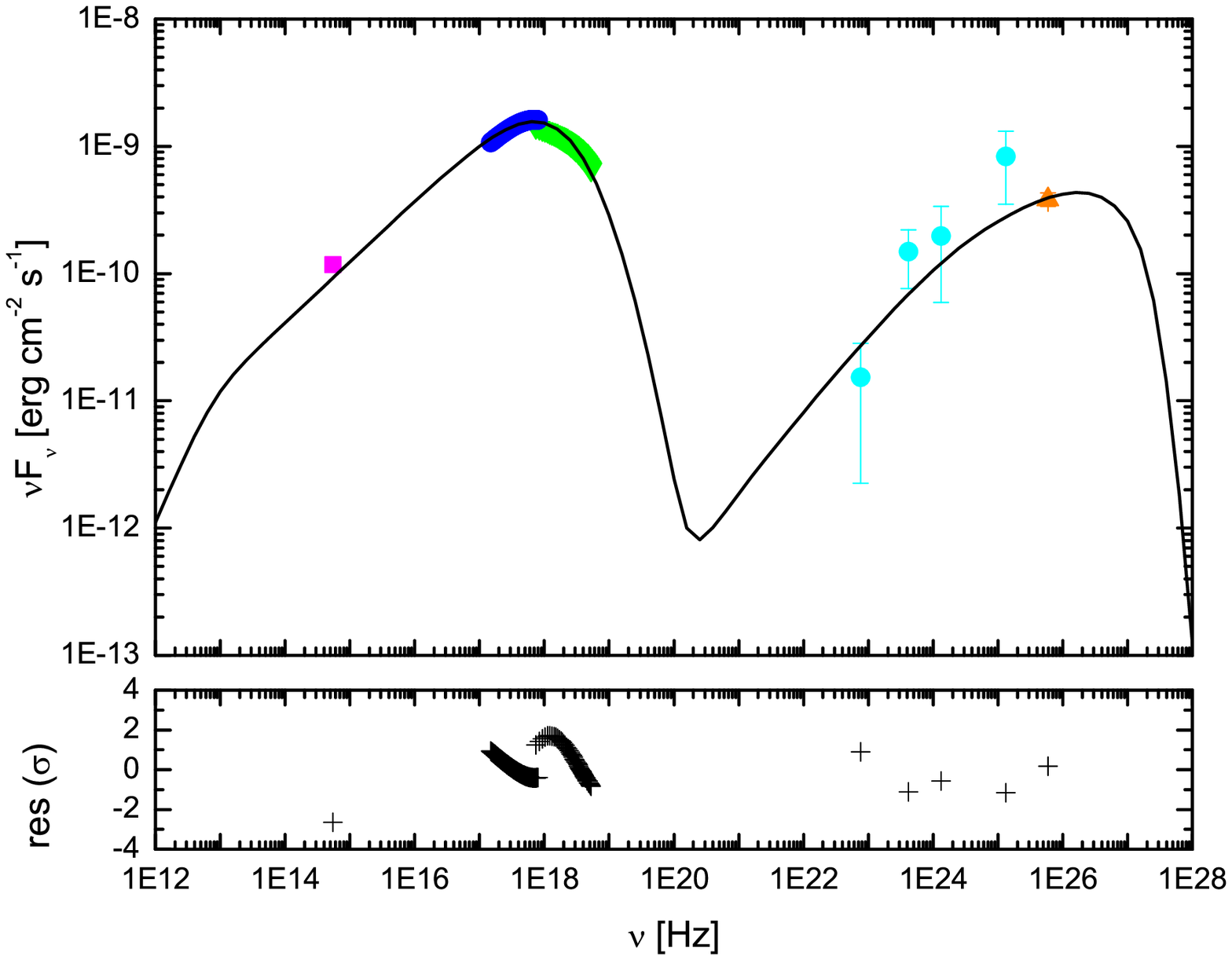}
\end{tabular}
  \end{center}
\caption{Modeling the SED in giant flare state with the LP
electron distribution. The curves are the same as that in
Fig.~\ref{fig:lowEc}, and the symbols are the same as that in
Fig.~\ref{fig:highEc}. } \label{fig:highlp}
\end{figure*}

Compared to the results in PLC case, it can be seen from
Fig.~\ref{fig:highlp} that the fittings with LP EED is
significantly improved with $\chi^2_{\nu}=0.30$ (for 317 d.o.f.).
Besides, relatively better constraints are obtained for all
parameters, especially for $s$, $r$, $\gamma^{\prime}_c$ and
$K^{\prime}_e$. Furthermore, the LP model with required $t_{\rm
v,min}\sim8$ hours can accommodate the intra-day/night variability
observed at GeV - TeV bands. It should be noted that the interpretations of the correlations of
parameters in Section~\ref{lowsed} still hold here.

\begin{deluxetable}{ccccccccc}
\tabletypesize{\scriptsize}
\tablecaption{Fitting parameters for the SED in giant flare state.}
\tablewidth{0pt}
\tablehead{
\colhead{Model} & \colhead{$\gamma^{\prime}_{c}$} & \colhead{$K^{\prime}_{e}$} &
\colhead{$s$} & \colhead{$r$} & \colhead{$B$} & \colhead{$t_{v,min}$} & \colhead{$\delta_{D}$} & \colhead{$U^{\prime}_e/U^{\prime}_B$}\\
 & \colhead{($10^5$)} & \colhead{($10^{53}$/$10^{42}$)}  & & & \colhead{(0.01 G)} &
 \colhead{($10^4$ s)} & \colhead{(10)} &
 }
\startdata
PLC model & 5.82$\pm$0.14  & 0.04$\pm$0.01  & 1.65$\pm$0.01  & -- & 1.07$\pm$0.06  & 7.98$\pm$0.55  & 2.99$\pm$0.08 & 110.08\\
  68\% limit & (5.66 - 5.97) & ( $<$0.04) & (1.64 - 1.66) & -- & ( $<$1.14) & (7.36 - 8.57) & ($<3.07$)\\
  \tableline
LP model & 3.63$\pm$0.21  & 0.17$\pm$0.03  & 2.02$\pm$0.04  & 3.79$\pm$0.26 & 2.67$\pm$0.32  & 2.66$\pm$0.32  & 3.44$\pm$0.20 & 61.75\\
  68\% limit & (3.42 - 3.86) & (0.14 - 0.20) & (1.99 - 2.06) & (3.51 - 4.08) & (2.30 - 3.03) & (2.30 - 3.02) & (3.22 - 3.66)\\
\enddata
\label{table:high}
\end{deluxetable}

\subsection{From low state to giant flare state}

After determining the EED in flare state, we can discuss the EED
in low state through investigating the variations of model
parameters with activities. \citet{Tramacere07,Tramacere09}
studied the spectral energy distributions (SEDs) of  Mrk 421 in
different states in the frame of the SSC model and found that
there is a negative correlation between the curvature parameter of
the radiation spectrum $b\approx r/5$ and $E_{\rm s}$, which is
expected from the stochastic acceleration mechanism.
\citet{Tramacere11} pointed out that the observed negative
correlation can be explained by the variation of the momentum
diffusion coefficient $D_p(\gamma)= D_{p0} (\gamma/\gamma_0)^q$,
where $q$ is the turbulence spectrum index (note that a larger
value of $D_{p}$ implies higher acceleration rate), or the fact
that the corresponding EEDs are far from the equilibrium where the
acceleration dominates over the radiative cooling. They also
suggested that the curvature increases as the radiative cooling
becomes important and EED is approaching to the equilibrium during
the evolution of EED. From Tables 1 and 2, if we assume that EEDs
in low and giant states both have the LP shapes, we find that
$\gamma'_c=1.47\times 10^5$, $B=2.05\times 10^{-2}$ G,
$\delta_{\rm D}=34.4$, and $r=1.68$ in the low state, and
$\gamma'_c=3.63\times 10^5$, $B=2.67\times 10^{-2}$ G,
$\delta_{\rm D}=33.4$, and $r=3.79$ in the giant state. Therefore,
the ratios of synchrotron peak energies and the curvature
parameters in two states are $[E_{\rm s}]_{\rm giant}/[E_{\rm
s}]_{\rm low}=[\gamma^{\prime 2}_{\rm c} B \delta_{\rm D}]_{\rm
giant}/[\gamma^{\prime 2}_{\rm c} B \delta_{\rm D}]_{\rm
low}\approx 7.7$ and $[r]_{\rm giant}/[r]_{\rm low}\approx2.3$,
which means that $E_{\rm s}$ increases with $r$. 
This scenario is not compatible with a purely accelerated-dominated 
transition, but the increase in the value of 
the curvature hints that during the high state,
the EED is at the equilibrium or very close, 
and that the cooling is dominating
over the acceleration. However, many other analyses 
\citep[e.g.,][]{becker,Stawarz,Tramacere11} 
pointed out that in the high state if the EED is close 
or at the equilibrium, the PLC EED would fit the SED better, 
while our results (Figs.~\ref{fig:highEc} and \ref{fig:highlp}) show that in high state the LP EED fits the SED better. 
Therefore, the above assumption of EEDs
in low and giant states both having the LP shapes is not correct.

Alternatively,  we still assume that the EEDs in low and giant
states both have the LP shapes and the different states are caused
by the variation of the momentum diffusion coefficient. From
Tables \ref{table:low} and \ref{table:high}, it can be found that
$\gamma^{\prime}_{c}$ and $r$ increases by a factor of $\sim 2$,
while $\delta_{\rm D}$ and $B$ almost keeps constant from low
state to flare state. Since $\gamma'_c$ is estimated by using the
condition $t_{\rm acc}(\gamma)=t_{\rm cool}(\gamma)$, where
$t_{\rm acc}(\gamma)\propto p^2/D_p\propto \gamma^{2-q}/D_{p0}$
\citep{Tramacere11} is the acceleration time and $t_{\rm
cool}(\gamma)\propto 1/\gamma$ (where the KN effect of IC
scattering is neglected) is the radiative cooling time, then we
have $\gamma'_c\propto D^{1/(3-q)}_{p0}$, and $\gamma'_c\propto
D_{p0}$ in the hard-sphere approximation ($q=2$). Hence,
$\gamma^{\prime}_{c}$ increases with $D_{\rm p0}$. On the other
hand, $r$ is inversely proportional to the momentum diffusion
coefficient, i. e., $r\propto D^{-1}_{\rm p0}$
\citep[e.g.,][]{Tramacere11,Massar011}. Therefore, the increase of
$D_{p0}$ cannot result in the increases of both $\gamma'_c$ and
$r$.

Since the SED in the giant state using the LP EED  can be fitted
better in comparison with that using PLC EED and  the
intra-day/night variability observed at GeV - TeV bands can be
accommodated in LP case (see \S 3.2 and Tables 1 and 2), therefore
the EED in the low state may be the PLC shape and the EED in the
giant flare state may be LP shape here.

\subsection{The EBL absorption}

\begin{figure*}[]
  \begin{center}
  \begin{tabular}{cc}
\hspace{-0.89cm}
     \includegraphics[width=90mm,height=80mm]{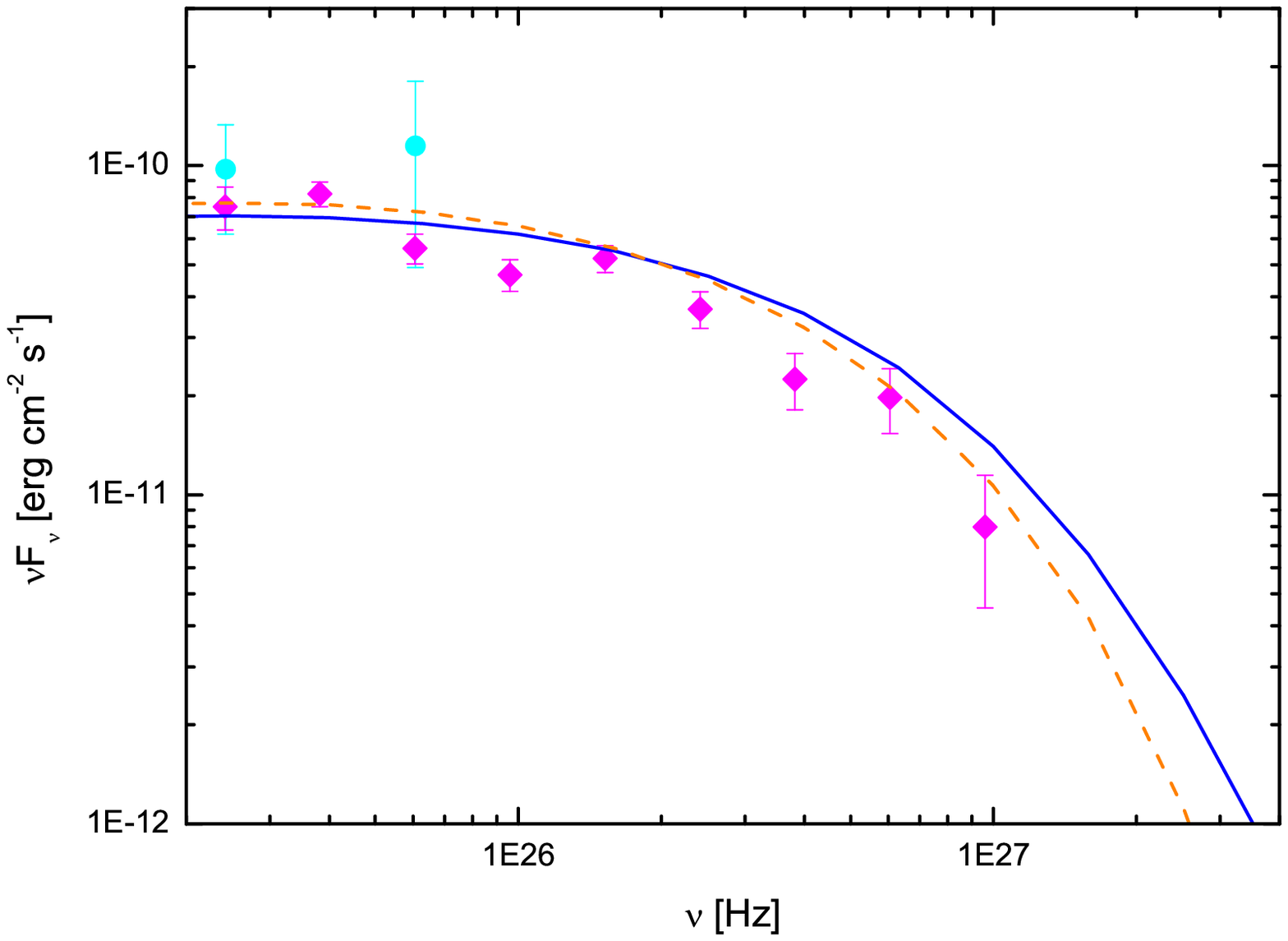} &
\hspace{-0.89cm}
     \includegraphics[width=90mm,height=80mm]{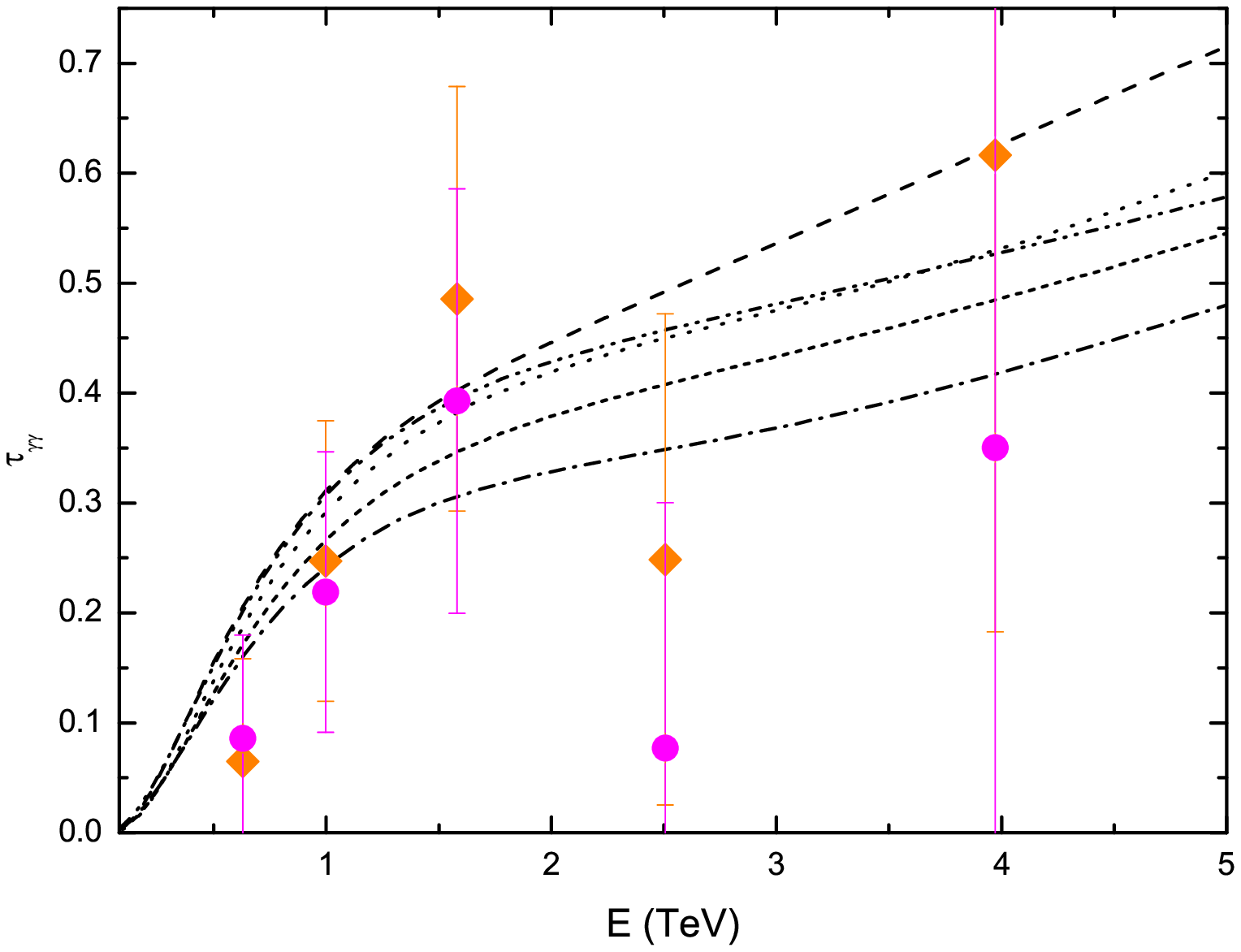}
\end{tabular}
  \end{center}
\caption{Left: Mrk 421 SED in low state in TeV window; The curves represent the TeV spectra calculated by LP model (solid line) and PLC model (dashed line). Right:  Optical depth $\tau_{\gamma\gamma}$; The symbols represent $\tau_{\gamma\gamma}$ derived by using TeV spectra calculated by LP model (diamonds) and PLC model (circles); The curves represent $\tau_{\gamma\gamma}$ calculated by EBL model of \citet{finke08} (dashed), \citet{fran08} (dotted), \citet{gil} (dash-double-dotted for WMAP5+fixed and short-dashed for fiducial WMAP5) and \citet{Kneiske} (dash-dotted).} \label{Fig:TeV}
\end{figure*}

In this Section, we investigate the EBL absorption. As an example
we take the SED in low state. Since the EBL absorption becomes
important when $E>\sim 2$ TeV for Mrk 421 ($z=0.031$), we compare
the TeV spectra predicted by both PLC and LP best-fit models,
which is shown in the left panle of Fig.~\ref{Fig:TeV}. It can be seen that the TeV
fluxes calculated by PLC and LP best-fit models are different when
E $>2$ TeV. Note that the predicted TeV flux is intrinsic flux,
therefore, the optical depth for EBL absorption on TeV photons
with energy $E$ is given by
\begin{equation}
\tau_{\gamma\gamma} (E)=\ln(f_{\rm int} (E)/f_{\rm obs} (E))\ ,
\end{equation}
where $f_{\rm int}$ is the TeV flux calculated by our best-fit
model from optical through GeV and $f_{\rm obs}$ is the measured
TeV flux. This optical depth can then be compared to the optical
depth calculated for the various EBL models
\citep[e.g.,][]{man10}.

We use the two extrapolated intrinsic TeV fluxes (see
Fig.~\ref{Fig:TeV}) to calculate $\tau_{\gamma\gamma}$.  The
results are shown in right panel of Fig.~\ref{Fig:TeV}.
Obviously, the values of
$\tau_{\gamma\gamma}$ calculated by using the TeV fluxes derived
in the LP model is almost two times of those in PLC model when E
$>2$ TeV. In the right panel of Fig~\ref{Fig:TeV}, for comparison, we also show
$\tau_{\gamma\gamma}$ calculated by three kinds of EBL models: the
high level one \citep[e.g.,][]{finke10}, the middle level one
\citep[e.g.,][]{fran08,gil} and the low level one
\citep[e.g.,][]{Kneiske}.  It can be seen that there are
discrepancies not only for the values of  $\tau_{\gamma\gamma}$
among various EBL models but also for those among each EBL model
and our results. Therefore, when such a method which a intrinsic
TeV spectrum is obtained from the extrapolation of the best-fit
spectrum from optical through GeV is used to constrain the EBL
models, the emission model of the source should be well determined
firstly, at least alternative model should be taken into account
to compare. Here, due to the large uncertainties of the measured
TeV data, we cannot obtain meaningful constraints on EBL models or
emission models.

As pointed out in \citet{Massaro3} and \citet{Tramacere09}, our
results distinctly show that the curvature in the TeV spectrum can
indeed affect the constraints on the EBL models. In our fitting,
we do not consider the data above 1TeV, however as discussed in
\citet{Tramacere11} the property of the curvature in the TeV
spectrum is more complex. Especially in the extreme KN regime, the
curvature of the IC emission is larger compared to the Thomson,
and it is almost equal to that of the EED \citep{Tramacere11}.
Therefore, the value of $r$ may be slightly underestimated in the
LP case although $r$ is mainly constrained by the synchrotron
emissions here, and this TeV flux may be overestimated.
Consequently, a subtle bias on $\tau_{\gamma\gamma}$ may be
introduced. Therefore, it is suggested that in order to constrain
the EBL more precisely, the complete TeV data should be taken into
account in the fitting. In this paper, due to our purpose and the
large errors of TeV data, our conclusion on the EBL constraint is
still robust.

\section{Discussion and Conclusions}
\label{DC}

Using the MCMC method, we study the SED of Mrk 421 in different
states for two kinds of EEDs with clear physical meanings: PLC and
LP EEDs. Our results indicate that
the EED in giant flare state is the LP shape and the stochastic
turbulence acceleration is dominant, while in low activity the
EED may be the PLC shape and the
shock acceleration may play a more important role.
This giant flare may be attributed to the
re-acceleration of these electrons with PLC shape by stochastic process.
Basically, we can understand this
process from low state to flare state in the scenario proposed by
\citet{V05} that electrons injected at the shock front, then are
accelerated at shock by Fermi I process and subsequently by the
stochastic process in the downstream region. Here, we specify this
scenario according to our results that the Fermi I process
accelerated electrons are continuously injected into the emitting
blob, and the SSC radiation from the steady EED (PLC
shape) in the emitting blob is responsible for the emission of Mrk 421 in low state;
Subsequently, the electrons with PLC shape are re-accelerated by
stochastic process, and the EED with a significant curvature at high energies
(LP) is formed, whose radiation contributes to the emission in
flare state. A requirement in the scenario is that the magnetic
field turbulence spectrum must be the case of the so-called
``hard-sphere" approximation with the spectrum index $q=2$ since for
cases of Kolmogorov turbulence ($q=5/3$) and Kraichna ($q=3/2$)
turbulence, acceleration efficiency depend on the electron energy
($\gamma^{\prime}$), so that the acceleration efficiency for electrons
with $\gamma^{\prime}\sim10^5$ is very low and the escape would be
dominant over the acceleration of electrons
\citep[e.g.,][]{becker}. This scenario including acceleration
process also can account for the spectra hardening in flaring
\citep[e.g.,][]{kirk}. This scenario can be examined and the
details of the flare can be studied in the time-dependent model.
We will study them in the model including acceleration
process \citep[e.g.,][]{Kusunose,Kata06,yan} in the coming work.

We notice that the jet of Mrk 421 appears to be particle dominated
(see $U^{\prime}_e/U^{\prime}_B$ in
Table~\ref{table:low},\ref{table:high}), which is consistent with
the result derived by \citet{man11}.
Finally, we stress the caveat on EBL constraints we derived.
The best-fit
spectrum from optical through GeV is often extrapolated into TeV
regime, and is considered as the intrinsic TeV spectrum. However,
our results show that with different emission models, the best-fit
SEDs from optical through GeV give different TeV spectra.
Therefore, it should be considered in caution when such
extrapolated TeV spectra is used to constrain the EBL models or
redshift of source. The accurately measured VHE spectrum at E $>2$
TeV bands \citep[e.g.,][]{Raue} could put more constraints on EBL
models and emission models.

\acknowledgments We thank David Paneque, Amit Shukla and Justin
Finke for sending us the observed data sets we used in this paper
and the anonymous referee for his/her very constructive comments.
We acknowledge the use of CosRayMC \citep{Liuj} adapted from the
COSMOMC package \citep{Lewis}. This work is partially supported by
the 973 Programme (2009CB824800) and by the Yunnan Province under
a grant 2009 OC. Q.Y. acknowledges the support of National Natural
Science Foundation of China under grant No. 11105155. Z.H.F.
acknowledges the support of National Natural Science Foundation of
China under grant No. 10963004.

\clearpage

\end{document}